\documentclass{article}
\usepackage{amssymb,amsmath}
\usepackage{epsfig}

\begin{document}
\title{Thermodynamical Cross-Relations for Superconductors at the Critical Point}
\author{B.Grabiec and M.Matlak\thanks{matlak@server.phys.us.edu.pl}\\
Institute of Physics, Silesian University,\\
4 Uniwersytecka, PL-40-007 Katowice, Poland}
\maketitle

\begin{abstract}
We investigate superconducting systems with the use of the phenomenological
Landau's theory of second order phase transitions, including into the
considerations the critical behaviour of the chemical potential. We derive
in this way a variety of new thermodynamical relations at the critical
point. Twelve basic relations connect critical jumps of different
thermodynamical quantites (specific heat, chemical potential derivatives
with respect to temperature, pressure (volume) and number of particles,
volume (pressure) derivatives with respect to temperature and pressure
(volume)) with the critical temperature or its derivatives with respect to
the number of particles or pressure (volume). These relations allow to find
plenty of cross-relations between different quantities at the critical
point. The derived formulae can practically be used in many cases to find
such thermodynamical quantities at the critical point which are extremely
difficult to measure under the assumption that the other ones are already
known. We additionally perform a test of the two derived relations by using
two-band microscopic model, describing superconducting systems. We calculate
the specific heat, order parameter and chemical potential as functions of
temperature to show that the tested relations are very well fulfilled.


\end{abstract}

\paragraph{Introduction}

First observations of the chemical potential critical behaviour at second
order phase transitions has been performed for superconductors (see Refs
[1-10]). It has been found that at the critical temperature $T_s$ (phase
transition: superconductor - normal system) the temperature dependence of
the chemical potential exhibits a kink (change in slope), manifesting in
this way the appearance of the phase transition (cf also Refs [11-23]).
For superconducting systems (see Ref. [6]), an interesting relation between
the jump of the chemical potential temperature derivative and the jump of
the specific heat at the critical point has been derived with the use of the
phenomenological, Landau's theory of second order phase transitions (cf Refs
[24, 25]). It is, however, possible to include in a more complete way the
chemical potential critical behaviour into the general scheme of the
Landau's theory for superconducting systems to derive new relations between
different thermodynamical quantities at the critical point. Starting from
the Gibbs potential we derive six basic relations and further six ones are
resulting from the Helmholtz potential. There are two relations connecting
the critical jump of the specific heat at constant pressure (volume) and the
critical temperature. The next two relate the critical jump of the chemical
potential temperature derivative and the critical temperature derivative
with respect to the number of particles (constant pressure (volume)).
Critical jumps of the volume (pressure) derivatives with respect to
temperature and pressure (volume) can be connected with the critical
temperature derivatives with respect to pressure (volume), resulting in four
relations. The next four relations connect critical jumps of the chemical
potential derivatives with respect to pressure (volume) and the number of
particles with the critical temperature derivatives with respect to the
number of particles and pressure (volume). These twelve basic relations can
be combined together by forming ratios of different critical jumps. In this
way we eliminate an unknown parameter of the Landau's theory (second
Landau's parameter) and we connect directly the mentioned ratios with the
critical temperature of the system and its derivatives. These relations can
be very useful in the practice because they allow to find an unknown
critical jump (or unknown critical temperature derivative) when the other
quantities are already known from experimental data. One can additionally
eliminate the critical temperature derivatives from these relations to
obtain six independent cross-relations between different critical jumps and
the critical temperature of the system. These formulae can also be
practically used to find an uknown critical jump under the assumption that
the other critical jumps are already known. We additionally perform a test
of two basic relations, desribed above, when independently calculate the
second Landau's parameter from these two relations. To this purpose we
applied two-band, microscopic model, describing superconducting systems. The
numerical calculation of the specific heat jump and the jump of the order
parameter at the critical point, as well as, the critical temperature
derivative with respect to the number of particles (electrons) reveals
almost the same value for the second Landau's parameter when calculate it
from these two independent relations. It seems to be very strange that the
analytical derivation of the formulae, we have numerically tested, is
extremely difficult to do for this particular, microscopic model (we could
not perform it) because of its great complexity (to many implicit equations
to solve) whereas the phenomenological derivation is not only general
(independent from the assumed microscopic model) and straightforward but
also true (!).

\paragraph{Phenomenological relations}
Within the phenomenological theory of second order phase transitions (see
Refs [24, 25]) one assumes the existence of the order parameter $x$.
For simplicity, we assume that the system can be described by only one order
parameter \footnote{See also Note added in proof}. For superconducting systems the order parameter $x=|\Delta _s|^2$
where $\Delta _s$ is the superconducting order parameter (or its amplitude
in the case of anisotropic superconductivity). The order parameter should
fulfil a selfconsistent equation of the general form

\begin{equation}
x=f(T,p,N,x)
\end{equation}
where the variables $T,$ $p,$ $N$ denote the absolute temperature, pressure
and number of particles of the system. This equation can formally be
resolved to give an equation

\begin{equation}
x=x(T,p,N)
\end{equation}
of such a property that $x=0$ for $T{\geqslant}T_{s}$ where $T_{s}=T_{s}(p,N)$ is
the critical temperature of the transition from superconducting to normal
phase. One assumes also that the thermodynamic properties of the system can
be desribed by the Gibbs potential
\begin{equation}
Z=Z(T,p,N,x).
\end{equation}
In the vicinity of $T=T_s$ ($T<T_s$) the order parameter $x$ is very small
and therefore we can expand the Gibbs potential (3) into the Taylor series
(we retain quadratic terms)
\begin{equation}
Z(T,p,N,x)=Z_0(T,p,N)+a(T,p,N)x+\frac 12b(T,p,N)x^2
\end{equation}
where $a$ and $b$ are first and second derivatives of $Z$ with respect to $x$
at the point $x=0$. The minimum principle ($\displaystyle{\frac{\partial Z}{%
\partial x}=0},$ $\displaystyle{\frac{{\partial }^2Z}{{\partial }{x}^2}>0)}$%
, aplied to (4) results in the expression
\begin{equation}
\begin{array}{ll}
x=-\displaystyle{\frac ab} & (b>0).
\end{array}
\end{equation}
To obtain a general form of the phenomenological parameter $a=a(T,p,N)$ in
the vicinity of the critical point we can apply again the Taylor expansion
with respect to a small quantity $(T-T_s)$. We get
\begin{eqnarray}
a(T,p,N) &=&a(T_s+(T-T_s),p,N)  \nonumber \\
&\approx &a(T_s,p,N)+(T-T_s)\left( \frac{\partial a}{\partial T}\right)
_{p,N,T=T_s}.
\end{eqnarray}
However, at $T=T_s$ the order parameter $x=0.$ It means also that $%
a(T_s,p,N)=0$ (see (6) and (5)). Thus, we obtain
\begin{equation}
a(T,p,N)=\left\{
\begin{array}{ll}
(T-T_s)\displaystyle{\left( \frac{\partial a}{\partial T}\right) _{p,N,T=T_s}%
} & ,T\leqslant T_s \\
&  \\
0 & ,T>T_s
\end{array}
\right.
\end{equation}
and
\begin{equation}
x=x(T,p,N)=\left\{
\begin{array}{ll}
-\displaystyle{\frac{(T-T_s)}b\left( \frac{\partial a}{\partial T}\right)
_{p,N,T=T_s}} & ,T\leqslant T_s \\
&  \\
0 & ,T>T_s
\end{array}
\right. .
\end{equation}
Eliminating the phenomenological parameter $b$ from (4) by means of (5) we
get
\begin{equation}
Z=Z_0+\frac 12ax
\end{equation}
and using the thermodynamic relations
\begin{equation}
S=-\left( \frac{\partial Z}{\partial T}\right) _{p,N},
\end{equation}

\begin{equation}
S_0=-\left( \frac{\partial Z_0}{\partial T}\right) _{p,N}
\end{equation}
we can write
\begin{equation}
S-S_0=-\frac 12\left( \frac{\partial a}{\partial T}\right) _{p,N}x-\frac
12a\left( \frac{\partial x}{\partial T}\right) _{p,N}.
\end{equation}
The specific heat difference, produced by (12) is given by relation

\begin{eqnarray}
c_{p,N}(T)-c_{p,N}^{(0)}(T) &=&T\left( \frac{\partial S}{\partial T}\right)
_{p,N}-T\left( \frac{\partial S_0}{\partial T}\right) _{p,N}  \nonumber \\
&=&-\frac T2\left( \frac{\partial ^2a}{\partial T^2}\right) _{p,N}x-T\left(
\frac{\partial a}{\partial T}\right) _{p,N}\left( \frac{\partial x}{\partial
T}\right) _{p,N}  \nonumber \\
&&-\frac{aT}2\left( \frac{\partial ^2x}{\partial T^2}\right) _{p,N}.
\end{eqnarray}
To obtain the jump of the specific heat $\Delta c_{p,N}$ at the critical
point we have to insert $T=T_s$ into (13) taking also into accout that $%
x=a=0 $ at $T=T_s$ (see (7), (8)). Thus, we obtain
\begin{equation}
\Delta c_{p,N}=-T_s\left( \frac{\partial a}{\partial T}\right)
_{p,N,T=T_s}\left( \frac{\partial x}{\partial T}\right) _{p,N,T=T_s}.
\end{equation}
However, when looking at (8) we can see that
\begin{equation}
\left( \frac{\partial x}{\partial T}\right) _{p,N,T=T_s}=-\frac 1b\left(
\frac{\partial a}{\partial T}\right) _{p,N,T=T_s}\equiv \Delta \left( \frac{%
\partial x}{\partial T}\right) _{p,N,T=T_s}
\end{equation}
where ${\Delta }\displaystyle{\left( \frac{\partial x}{\partial T}\right) }%
_{p,N,T=T_s}$ is just the jump of the temperature derivative of the order
parameter at $T=T_s$ because the order parameter $x$ vanishes for $%
T\geqslant T_s$. Thus, we can rewrite (14) with the use of (15) by
eliminating $\displaystyle{\left( \frac{\partial a}{\partial T}\right) }%
_{p,N,T=T_s}$. We obtain
\begin{equation}
\Delta c_{p,N}=bT_s\left[ \Delta \left( \frac{\partial x}{\partial T}\right)
_{p,N,T=T_s}\right] ^2.
\end{equation}

The chemical potential of the system can be introduced from the general
expression for the Gibbs potential using the relations
\begin{equation}
\begin{array}{lll}
\displaystyle{\mu =\left( \frac{\partial Z}{\partial N}\right) _{T,p}} & , & %
\displaystyle{\mu _0=\left( \frac{\partial Z_0}{\partial N}\right) _{T,p}.}
\end{array}
\end{equation}
These relations applied to (9) result
\begin{equation}
\mu =\mu _0+\frac 12\left( \frac{\partial a}{\partial N}\right)
_{T,p}x+\frac 12a\left( \frac{\partial x}{\partial N}\right) _{T,p}.
\end{equation}
Thus, the jump of the chemical potential temperature derivative at $T=T_s$ $%
(x=0,$ $a=0$ for $T\geqslant T_s)$ is equal to
\begin{eqnarray}
\Delta \left( \frac{\partial \mu }{\partial T}\right) _{p,N,T=T_s} &\equiv
&\left( \frac{\partial \mu }{\partial T}\right) _{p,N,T=T_s}-\left( \frac{%
\partial \mu _0}{\partial T}\right) _{p,N,T=T_s}  \nonumber \\
&=&\frac 12\left( \frac{\partial a}{\partial N}\right) _{T=T_s,p}\left(
\frac{\partial x}{\partial T}\right) _{p,N,T=T_s} \\
&&+\frac 12\left( \frac{\partial a}{\partial T}\right) _{p,N,T=T_s}\left(
\frac{\partial x}{\partial N}\right) _{T=T_s,p}.  \nonumber
\end{eqnarray}
Using (7), (8) and (15) we find $(T_s=T_s(p,N)):$%
\begin{eqnarray}
\left( \frac{\partial a}{\partial N}\right) _{T=T_s,p} &=&-\left( \frac{%
\partial T_s}{\partial N}\right) _p\left( \frac{\partial a}{\partial T}%
\right) _{p,N,T=T_s}  \nonumber \\[0.25cm]
&=&b\left( \frac{\partial T_s}{\partial N}\right) _p\left[ \Delta \left(
\frac{\partial x}{\partial T}\right) _{p,N,T=T_s}\right]
\end{eqnarray}
and

\begin{eqnarray}
\left( \frac{\partial x}{\partial N}\right) _{T=T_s,p} &=&\frac 1b\left(
\frac{\partial T_s}{\partial N}\right) _p\left( \frac{\partial a}{\partial T}%
\right) _{p,N,T=T_s}  \nonumber \\[0.25cm]
&=&-\left( \frac{\partial T_s}{\partial N}\right) _p\left[ \Delta \left(
\frac{\partial x}{\partial T}\right) _{p,N,T=T_s}\right] .
\end{eqnarray}
Introducing (20) and (21) into (19) we get
\begin{equation}
\Delta \displaystyle{\left( \frac{\partial \mu }{\partial T}\right)
_{p,N,T=T_s}}=b\displaystyle{\left( \frac{\partial T_s}{\partial N}\right)
_p\left[ \Delta \left( \frac{\partial x}{\partial T}\right)
_{p,N,T=T_s}\right] ^2.}
\end{equation}
It is also easy to find that

\begin{equation}
\left( \frac{\partial a}{\partial p}\right) _{T=T_s,N}=-\left( \frac{%
\partial T_s}{\partial p}\right) _N\left( \frac{\partial a}{\partial T}%
\right) _{p,N,T=T_s}=b\left( \frac{\partial T_s}{\partial p}\right) _N{%
\Delta }\left( \frac{\partial x}{\partial T}\right) _{p,N,T=T_s},
\end{equation}

\begin{equation}
\left( \frac{\partial x}{\partial p}\right) _{T=T_s,N}=\frac 1b\left( \frac{%
\partial T_s}{\partial p}\right) _N\left( \frac{\partial a}{\partial T}%
\right) _{p,N,T=T_s}=-\left( \frac{\partial T_s}{\partial p}\right) _N{%
\Delta }\left( \frac{\partial x}{\partial T}\right) _{p,N,T=T_s}.
\end{equation}
We can perform similar calculations as above to find critical jumps of the
following thermodynamical quantities: ${\Delta }\left( \frac{\partial V}{%
\partial T}\right) _{p,N,T=T_s}$, ${\Delta }\left( \frac{\partial \mu }{%
\partial p}\right) _{T=T_s,N}$, ${\Delta }\left( \frac{\partial \mu }{%
\partial N}\right) _{T=T_s,p}$ and ${\Delta }\left( \frac{\partial V}{%
\partial p}\right) _{T=T_s,N}$ . Starting from the expression for the Gibbs
potential (9) with the use of the thermodynamical relations ${\mu }=\left(
\frac{\partial Z}{\partial T}\right) _{T,p}$ (${\mu }_0=\left( \frac{%
\partial Z_0}{\partial T}\right) _{T,p}$), $V=\left( \frac{\partial Z}{%
\partial p}\right) _{T,N}$ ($V_0=\left( \frac{\partial Z_0}{\partial p}%
\right) _{T,N}$) and the above formulae (15), (20), (21), (23) and (24) we
obtain

\begin{equation}
{\Delta }\left( \frac{\partial V}{\partial T}\right) _{p,N,T=T_s}=b\left(
\frac{\partial T_s}{\partial p}\right) _N{\left[ \Delta \left( \frac{%
\partial x}{\partial T}\right) _{p,N,T=T_s}\right] ^2},
\end{equation}

\begin{equation}
{\Delta }\left( \frac{\partial \mu }{\partial p}\right) _{T=T_s,N}=-b\left(
\frac{\partial T_s}{\partial N}\right) _p\left( \frac{\partial T_s}{\partial
p}\right) _N{\left[ \Delta \left( \frac{\partial x}{\partial T}\right)
_{p,N,T=T_s}\right] ^2},
\end{equation}

\begin{equation}
{\Delta }\left( \frac{\partial \mu }{\partial N}\right) _{T=T_s,p}=-b\left[
\left( \frac{\partial T_s}{\partial N}\right) _p\right] ^2{\left[ \Delta
\left( \frac{\partial x}{\partial T}\right) _{p,N,T=T_s}\right] ^2}
\end{equation}
and
\begin{equation}
{\Delta }\left( \frac{\partial V}{\partial p}\right) _{T=T_s,N}=-b\left[
\left( \frac{\partial T_s}{\partial p}\right) _N\right] ^2{\left[ \Delta
\left( \frac{\partial x}{\partial T}\right) _{p,N,T=T_s}\right] ^2}.
\end{equation}
Using the Maxwell relations $\left( \frac{\partial S}{\partial p}\right)
_{T,N}=-\left( \frac{\partial V}{\partial T}\right) _{p,N}$, $\left( \frac{%
\partial S}{\partial N}\right) _{T,p}=-\left( \frac{\partial \mu }{\partial T%
}\right) _{p,N}$ and $\left( \frac{\partial V}{\partial N}\right)
_{T,p}$ $=\left( \frac{\partial \mu }{\partial p}\right) _{T,N}$ we can see
that the other critical jumps $\Delta \left( \frac{\partial S}{\partial p}\right) _{T=T_s,N}$
, $\Delta \left( \frac{\partial S}{\partial N}\right)
_{T=T_s,p}$ and $\Delta \left( \frac{\partial V}{\partial N}\right)
_{T=T_s,p}$ can easily be obtained from (25), (22) and (26), respectively.
In other words, there are only six independent expressions describing the
critical jumps of the first derivatives ((16), (22) and (25)-(28)). These
expressions are very important because with the use of them we can find many
interesting cross-relations. Dividing e.g. (22) and (25)-(28) by (16) we
obtain five independent ratios (the specific heat critical jump is finite
for superconductors):

\begin{equation}
\frac{\Delta \left( \frac{\partial \mu }{\partial T}\right) _{p,N,T=T_s}}{%
\Delta c_{p,N}}=\left( \frac{\partial \ln T_s}{\partial N}\right) _p,
\end{equation}

\begin{equation}
\frac{\Delta \left( \frac{\partial V}{\partial T}\right) _{p,N,T=T_s}}{%
\Delta c_{p,N}}=\left( \frac{\partial \ln T_s}{\partial p}\right) _N,
\end{equation}

\begin{equation}
\frac{\Delta \left( \frac{\partial \mu }{\partial p}\right) _{T=T_s,N}}{%
\Delta c_{p,N}}=-\left( \frac{\partial \ln T_s}{\partial N}\right) _p\left(
\frac{\partial T_s}{\partial p}\right) _N,
\end{equation}

\begin{equation}
\frac{\Delta \left( \frac{\partial \mu }{\partial N}\right) _{T=T_s,p}}{%
\Delta c_{p,N}}=-\left( \frac{\partial \ln T_s}{\partial N}\right) _p\left(
\frac{\partial T_s}{\partial N}\right) _p,
\end{equation}
and
\begin{equation}
\frac{\Delta \left( \frac{\partial V}{\partial p}\right) _{T=T_s,N}}{\Delta
c_{p,N}}=-\left( \frac{\partial \ln T_s}{\partial p}\right) _N\left( \frac{%
\partial T_s}{\partial p}\right) _N.
\end{equation}
It is, however, possible to find also another ratios (equivalent to the
above relations). When eliminate the derivatives $\left( \frac{\partial T_s}{%
\partial N}\right) _p$ and $\left( \frac{\partial T_s}{\partial p}\right) _N$
from the above expressions we can obtain three independent cross-relations
connecting different critical jumps: 
\begin{equation}
\Delta \left( \frac{\partial \mu }{\partial p}\right) _{T=T_s,N}\cdot \Delta
c_{p,N}=-T_s\Delta \left( \frac{\partial \mu }{\partial T}\right)
_{p,N,T=T_s}\cdot \Delta \left( \frac{\partial V}{\partial T}\right)
_{p,N,T=T_s},
\end{equation}

\begin{equation}
\Delta \left( \frac{\partial \mu }{\partial N}\right) _{T=T_s,p}\cdot \Delta
c_{p,N}=-T_s\left[ \Delta \left( \frac{\partial \mu }{\partial T}\right)
_{p,N,T=T_s}\right] ^2
\end{equation}
and
\begin{equation}
\Delta \left( \frac{\partial V}{\partial p}\right) _{T=T_s,N}\cdot \Delta
c_{p,N}=-T_s\left[ \Delta \left( \frac{\partial V}{\partial T}\right)
_{p,N,T=T_s}\right] ^2.
\end{equation}
These relations allow to find some critical jumps which are difficult to
measure, knowing the other from experiments, without necessity to know the
derivatives of the critical temperature.

We can also derive several additional relations, when using the
thermodynamical relationship resulting from the Gibbs potential (1)
\begin{equation}
V=\left( \frac{\partial Z}{\partial p}\right) _{T,N}=V(T,p,N,x).
\end{equation}
The equation (37) can formally be resolved to obtain the equation of state
\begin{equation}
p=p(T,V,N,x).
\end{equation}
We can insert it into (1) to obtain
\begin{equation}
{\bar{Z}}={\bar{Z}}(T,V,N,x)
\end{equation}
which is nothing else but the Helmholtz potential. We can further apply
exactly the same procedure as described above starting again with the Taylor
expansion of the expression (39) with respect to the order parameter in the
vicinity of the critical point. Therefore there is no need to demonstrate
here all the details of the performed calculations. We restrict ourselves to
present only final results. The analogous formulae (cf (16), (22) and
(25)-(28)) for the critical jumps, resulting from (39), have the form

\begin{equation}
{\Delta }c_{V,N}={b}T_s{\left[ \Delta \left( \frac{\partial x}{\partial T}%
\right) _{V,N,T=T_s}\right] }^2,
\end{equation}

\begin{equation}
{\Delta }\left( \frac{\partial \mu }{\partial T}\right) _{V,N,T=T_s}={b}%
\left( \frac{\partial T_s}{\partial N}\right) _V{\left[ \Delta \left( \frac{%
\partial x}{\partial T}\right) _{V,N,T=T_s}\right] ^2},
\end{equation}

\begin{equation}
{\Delta }\left( \frac{\partial p}{\partial T}\right) _{V,N,T=T_s}=-{b}\left(
\frac{\partial T_s}{\partial V}\right) _N{\left[ \Delta \left( \frac{%
\partial x}{\partial T}\right) _{V,N,T=T_s}\right] ^2},
\end{equation}

\begin{equation}
{\Delta }\left( \frac{\partial \mu }{\partial V}\right) _{T=T_s,N}=-{b}%
\left( \frac{\partial T_s}{\partial N}\right) _V\left( \frac{\partial T_s}{%
\partial V}\right) _N{\left[ \Delta \left( \frac{\partial x}{\partial T}%
\right) _{V,N,T=T_s}\right] ^2},
\end{equation}

\begin{equation}
{\Delta }\left( \frac{\partial \mu }{\partial N}\right) _{T=T_s,V}=-{b}%
\left[ \left( \frac{\partial T_s}{\partial N}\right) _V\right] ^2{\left[
\Delta \left( \frac{\partial x}{\partial T}\right) _{V,N,T=T_s}\right] ^2}
\end{equation}
and

\begin{equation}
{\Delta }\left( \frac{\partial p}{\partial V}\right) _{T=T_s,N}={b}\left[
\left( \frac{\partial T_s}{\partial V}\right) _N\right] ^2{\left[ \Delta
\left( \frac{\partial x}{\partial T}\right) _{V,N,T=T_s}\right] ^2}.
\end{equation}
The constant $b$ has the same meaning as before but actually $b=b(T_s,V,N)$.
To derive (40)-(45) we have applied thermodynamical relations $S=-\left(
\frac{\partial {\bar{Z}}}{\partial T}\right) _{V,N}$, $p=-\left( \frac{%
\partial {\bar{Z}}}{\partial V}\right) _{T,N}$ and $\mu =\left( \frac{%
\partial {\bar{Z}}}{\partial N}\right) _{T,V}$ ($S_0=-\left( \frac{\partial {%
\bar{Z}}_0}{\partial T}\right) _{V,N}$, $p_0=-\left( \frac{\partial {\bar{Z}}%
_0}{\partial V}\right) _{T,N}$ and ${\mu }_0=\left( \frac{\partial {\bar{Z}}%
_0}{\partial N}\right) _{T,V}$ where ${\bar{Z}}_0={\bar{Z}}_0(T,V,N)$ in
analogy to (4)). Because of the Maxwell relations $\left( \frac{\partial S}{%
\partial V}\right) _{T,N}=\left( \frac{\partial p}{\partial T}\right) _{V,N}$%
, $\left( \frac{\partial p}{\partial N}\right) _{T,V}=-\left( \frac{\partial
\mu }{\partial V}\right) _{T,N}$ and $\left( \frac{\partial S}{\partial N}%
\right) _{T,V}=-\left( \frac{\partial \mu }{\partial T}\right) _{V,N}$ the
critical jumps ${\Delta }\left( \frac{\partial S}{\partial V}\right)
_{T=T_s,N},$ ${\Delta }\left( \frac{\partial p}{\partial N}\right)
_{T=T_s,V} $ and ${\Delta }\left( \frac{\partial S}{\partial N}\right)
_{T=T_s,V}$ can easily be found from (42), (43) and (41), respectively,
therefore there are again only six independent formulae (40)-(45). When e.g.
devide (41)-(45) by (40) we get:

\begin{equation}
\frac{\Delta \left( \frac{\partial \mu }{\partial T}\right) _{V,N,T=T_s}}{%
\Delta c_{V,N}}=\left( \frac{\partial \ln T_s}{\partial N}\right) _V,
\end{equation}

\begin{equation}
\frac{\Delta \left( \frac{\partial p}{\partial T}\right) _{V,N,T=T_s}}{%
\Delta c_{V,N}}=-\left( \frac{\partial \ln T_s}{\partial V}\right) _N,
\end{equation}

\begin{equation}
\frac{\Delta \left( \frac{\partial \mu }{\partial V}\right) _{T=T_s,N}}{%
\Delta c_{V,N}}=-\left( \frac{\partial \ln T_s}{\partial N}\right) _V\left(
\frac{\partial T_s}{\partial V}\right) _N,
\end{equation}

\begin{equation}
\frac{\Delta \left( \frac{\partial \mu }{\partial N}\right) _{T=T_s,V}}{%
\Delta c_{V,N}}=-\left( \frac{\partial \ln T_s}{\partial N}\right) _V\left(
\frac{\partial T_s}{\partial N}\right) _V
\end{equation}
and

\begin{equation}
\frac{\Delta \left( \frac{\partial p}{\partial V}\right) _{T=T_s,N}}{\Delta
c_{V,N}}=\left( \frac{\partial \ln T_s}{\partial V}\right) _N\left( \frac{%
\partial T_s}{\partial V}\right) _N.
\end{equation}
There are, however, many other possilities to form the ratios, leading to
equivalent relations.

At last, when eliminate from (46)-(50) the critical temperature derivatives,
we obtain (cf (34)-(36)) 

\begin{equation}
\Delta \left( \frac{\partial \mu }{\partial V}\right) _{T=T_s,N}\cdot \Delta
c_{V,N}=T_s\Delta \left( \frac{\partial \mu }{\partial T}\right)
_{V,N,T=T_s}\cdot \Delta \left( \frac{\partial p}{\partial T}\right)
_{V,N,T=T_s},
\end{equation}

\begin{equation}
\Delta \left( \frac{\partial \mu }{\partial N}\right) _{T=T_s,V}\cdot \Delta
c_{V,N}=-T_s\left[ \Delta \left( \frac{\partial \mu }{\partial T}\right)
_{V,N,T=T_s}\right] ^2
\end{equation}
and

\begin{equation}
\Delta \left( \frac{\partial p}{\partial V}\right) _{T=T_s,N}\cdot \Delta
c_{V,N}=T_s\left[ \Delta \left( \frac{\partial p}{\partial T}\right)
_{V,N,T=T_s}\right] ^2.
\end{equation}

Thus, the derived formulae (16), (22), (25)-(28) together with (40)-(45)
allowed to find many useful thermodynamical cross-relations for
superconducting systems at the critical point.

\paragraph{Numerical Test: superconductor - normal system transition}

In the following we perform a validity test of the formulae (40) and (41)
when using microscopic model, describing superconducting properties of a
system with two hybrydized electronic bands. The Cooper - pairs attraction,
responsible for the superconducting properties, we choose in the form of the
intersite interaction restricted to only one band (one superconducting order
parameter). The Hamiltonian of this model can be written in the form (great
canonical ensemble)

\begin{equation}
\overline{H}=H-{\mu }N
\end{equation}
where

\begin{eqnarray}
H &=&\displaystyle{\sum\limits_{{\alpha }=1}^2\sum\limits_{i,j,{\sigma }}}%
t_{i,j}^{({\alpha })}c_{i,{\sigma }}^{({\alpha })^{+}}c_{i,{\sigma }}^{({%
\alpha })}+V\displaystyle{\sum\limits_{i,{\sigma }}}\left( c_{i,{\sigma }%
}^{(1)^{+}}c_{i,{\sigma }}^{(2)}+c_{i,{\sigma }}^{(2)^{+}}c_{i,{\sigma }%
}^{(1)}\right)  \nonumber \\
&&+\displaystyle{\sum\limits_{i,j}R_{i,j}}c_{i,{\uparrow }}^{(2)^{+}}c_{j,{%
\downarrow }}^{(2)^{+}}c_{j,{\downarrow }}^{(2)}c_{i,{\uparrow }}^{(2)}
\end{eqnarray}
and

\begin{equation}
N=\sum\limits_{\alpha =1}^2\sum\limits_{i,\sigma }n_{i,\sigma }^{(\alpha )}.
\end{equation}
Two electronic bands are characterized by the hopping intergrals $t_{i,j}^{({%
\alpha })}$ (${\alpha }=1,2$; $t_{i,i}^{(1)}=0$, $t_{i,i}^{(2)}=t$). The
parameters $V$ and $R_{i,j}$ denote the hybridization and intersite Cooper -
pairs attraction, respectively. The electronic annihilation (creation)
operators (${\alpha }=1,2$; ${\sigma }={\uparrow },{\downarrow }$) are
denoted by $c_{i,{\sigma }}^{({\alpha })}$ $(c_{i,{\sigma }}^{({\alpha }%
)^{+}})$, respectively, where $i$ is the lattice site index. For simplicity,
we consider simple cubic (sc) lattice. Similarly to Ref. [26] we apply the
mean field approximation (see the third term in (55))

\begin{eqnarray}
c_{i,{\uparrow }}^{(2)^{+}}c_{j,{\downarrow }}^{(2)^{+}}c_{j,{\downarrow }%
}^{(2)}c_{i,{\uparrow }}^{(2)} &\approx &\left\langle c_{i,{\uparrow }%
}^{(2)^{+}}c_{j,{\downarrow }}^{(2)^{+}}\right\rangle c_{j,{\downarrow }%
}^{(2)}c_{i,{\uparrow }}^{(2)}+\left\langle c_{j,{\downarrow }}^{(2)}c_{i,{%
\uparrow }}^{(2)}\right\rangle c_{i,{\uparrow }}^{(2)^{+}}c_{j,{\downarrow }%
}^{(2)^{+}}  \nonumber \\
&&-\left\langle c_{i,{\uparrow }}^{(2)^{+}}c_{j,{\downarrow }%
}^{(2)^{+}}\right\rangle \left\langle c_{j,{\downarrow }}^{(2)}c_{i,{%
\uparrow }}^{(2)}\right\rangle .
\end{eqnarray}
Introducing (57) into (55) and applying Fourier transformation we obtain

\begin{eqnarray}
H &=&-\displaystyle{\sum\limits_{{\bf k}}}{\Delta }_{{\bf k}}^{*}\langle c_{-%
{\bf {k}{\downarrow }}}^{(2)}c_{{\bf {k}{\uparrow }}}^{(2)}\rangle +%
\displaystyle{\sum\limits_{\alpha =1}^2\sum\limits_{{\bf k}}}{\varepsilon }_{%
{\bf k}}^{(\alpha )}(n_{{\bf {k},{\uparrow }}}^{(\alpha )}+n_{-{\bf {k},{%
\downarrow }}}^{(\alpha )})  \nonumber \\
&&+V\displaystyle{\sum\limits_{{\bf k}}}\left( c_{{\bf {k},{\uparrow }}%
}^{(1)^{+}}c_{{\bf {k},{\uparrow }}}^{(2)}+c_{{\bf {k},{\uparrow }}%
}^{(2)^{+}}c_{{\bf {k},{\uparrow }}}^{(1)}+c_{-{\bf {k},{\downarrow }}%
}^{(1)^{+}}c_{-{\bf {k},{\downarrow }}}^{(2)}+c_{-{\bf {k},{\downarrow }}%
}^{(2)^{+}}c_{-{\bf {k},{\downarrow }}}^{(1)}\right)  \nonumber \\
&&+\displaystyle{\sum\limits_{{\bf k}}}{\Delta }_{{\bf k}}^{*}c_{-{\bf {k},{%
\downarrow }}}^{(2)}c_{{\bf {k},{\uparrow }}}^{(2)}+\displaystyle{%
\sum\limits_{{\bf k}}}{\Delta }_{{\bf k}}c_{{\bf {k},{\uparrow }}}^{(2)}c_{-%
{\bf {k},{\downarrow }}}^{(2)},
\end{eqnarray}

\begin{equation}
N=\sum\limits_{\alpha =1}^2\sum\limits_{{\bf k}}(n_{{\bf {k},{\uparrow }}%
}^{(\alpha )}+n_{-{\bf {k},{\downarrow }}}^{(\alpha )})
\end{equation}
where

\begin{equation}
{\Delta }_{{\bf k}}=\displaystyle{\frac 1{{\cal N}}\sum\limits_{{\bf k}}}R(%
{\bf {k}-{k^{\prime }})\langle c_{-{k},{\downarrow }}^{(2)}c_{{k},{\uparrow }%
}^{(2)}\rangle ,}
\end{equation}

\begin{equation}
{\varepsilon}_{{\bf k}}^{(1)}=-\displaystyle{\frac{W^{(1)}}{6}} F({\bf k),}
\end{equation}

\begin{equation}
{\varepsilon }_{{\bf k}}^{(2)}=t-\displaystyle{\frac{W^{(2)}}6}F({\bf k),}
\end{equation}

\begin{equation}
R({\bf {k})}=2R_0F({\bf k),}
\end{equation}

\begin{equation}
F({\bf {k})}={\bf \cos (}k_{\text{x}}a{\bf )+\cos (}k_{\text{y}}a{\bf )+\cos
(}k_{\text{z}}a{\bf )}
\end{equation}
and ${\cal N}$ is the number of lattice points. We assume that the bandwidth
$W^{(2)}=W$ and $W^{(1)}={\delta }W$ $({\delta }\ll 1)$. In the following we
assume that (cf Ref. [26])

\begin{equation}
{\Delta }_{{\bf k}}={\Delta }_sF({\bf k)}
\end{equation}
where ${\Delta }_s$ (we assume ${\Delta }_s^{*}={\Delta }_s$) is the
amplitude of the superconducting order parameter ($s^{*}$ - symmetry).
Looking at the expressions (61)-(63) and (65) we see that the only
dispression in our system is generated by the formfactor (64) of the sc
lattice. It means that we can convert $\frac 1{{\cal N}}\sum\limits_{{\bf k}%
}(...){\rightarrow }\int\limits_{-\frac W2}^{\frac W2}{\rho }_0({\varepsilon
})(...)d{\varepsilon }$ where ${\rho }_0({\varepsilon })$ is the true
density of states of the sc lattice given by the Jelitto formula [27].
Applying the standard Green's function approach (cf e.g. [28]) we can find a
set of transcendental equations of the type (${\alpha }=1,2$; ${\sigma }={%
\uparrow },{\downarrow }$)

\begin{equation}
\langle n_{\sigma} ^{(\alpha )}\rangle =f_{\sigma}^{(\alpha )}(T,\Delta
_s^2,\mu ),
\end{equation}

\begin{equation}
{\Delta }_s^2={\Delta }_s^2f(T,{\Delta }_s^2,\mu )
\end{equation}
and

\begin{equation}
\displaystyle{\sum\limits_{\alpha =1}^2\sum\limits_\sigma }\langle n_\sigma
^{(\alpha )}\rangle =n
\end{equation}
where $\displaystyle{n=\frac{\langle N\rangle }{{\cal N}}}$ is the average
number of electrons per lattice site. The calculations of all necessary
quantities to check the formulae (40) and (41) are very laborious and lead
to very complicated expressions (to long to present here). Therefore we are
forced to describe only the general calculation scheme and to present final,
numerical results. The implicit equations (66)-(68) allow to calculate the
temperature dependence of the square amplitude ${\Delta }_s^2$ (it is just
the order parameter in the sense of the Landau's theory) and ${\mu }$ as
function of temperature. The calculation of the specific heat per lattice
site $c_{V,n}$ can be performed with the use of (58) when calculating

\begin{equation}
\widetilde{\left\langle H\right\rangle }=\frac{\langle H\rangle }{{\cal N}}
\end{equation}
and differentiating it with respect to $T$. This differentiation is,
however, very laborious because ${\Delta }_s^2$ and ${\mu }$ fulfil the
equations (66)-(68) and are also temperature dependent. In other words, we
have to differentiate them with respect to $T$ to calculate the temeprature
derivatives of ${\Delta }_s^2$ and ${\mu }$ (necessary to check the formulae
(40) and (41). The transition temperature $T_s$ (phase transition from
superconducting to normal state) can be obtained from the equation (67). It
is easy to see that $T_s$ should fulfil the following equation

\begin{equation}
f(T_s,{\mu })=1.
\end{equation}

\begin{figure}[t]
\epsfig{file=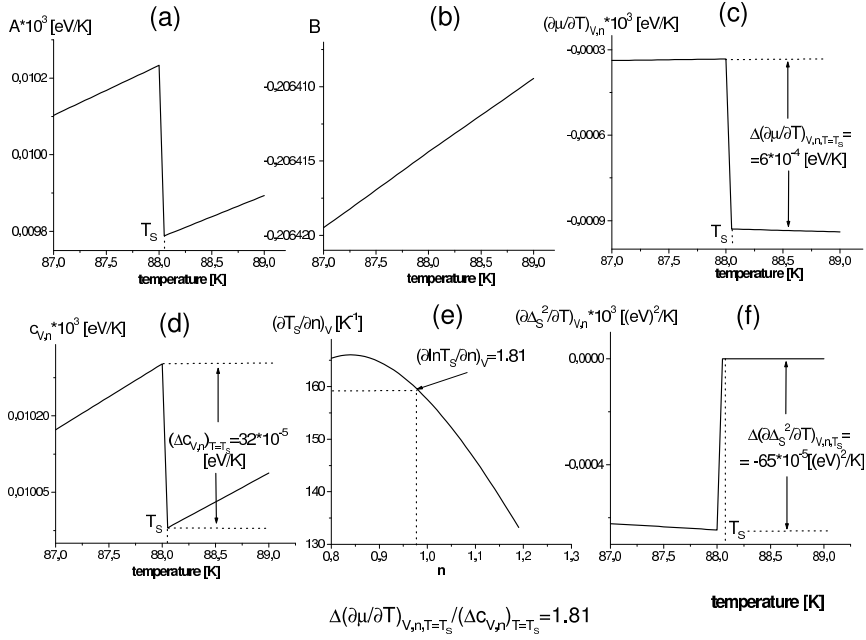,angle=-90,scale=0.45}
\caption{\footnotesize {Plot of the coefficients $A$
(Fig. 1a), $B$ (Fig. 1b) and ${\left( \frac{\partial
\mu }{\partial T}\right) }_{V,n}$(Fig. 1c) entering into the specific heat
per lattice site (Fig. 1d) as functions of temperature in the vicinity of
the critical point. The plots in Fig. 1e, f show the superconducting
critical temperature derivative with respect to the average number of
electrons per lattice site $n$ as function of $n$ and the temperature
derivative of the order parameter $\Delta _s^2$ as function of temperature.
The model parameters: $W$= 3 eV, $\delta $=0.25, $R_0$=0.6 eV, $t$=1 eV, $V$%
=0.15 eV, $n$=0.97.}}
\end{figure}

The function $f$ in (70) is exactly the same as in the rhs of (67) but we
have additionally assumed that the order parameter is equal to zero ($T=T_s$%
). We should also calculate the derivative $\displaystyle{\left( \frac{%
\partial {T_s}}{\partial {n}}\right) _V}$. It can be performed by the
differentiation of the equations (70), (66) and (68) with respect to $n$ (we
put $T=T_s$ therein (the order parameter is equal to zero)). In such a way
we have all necessary quantities to check the formulae (40) and (41). These
formulae we should first rewrite in the following way ($N\longrightarrow
\langle N\rangle ={\cal N}n$, $x={\Delta }_s^2,$ $b\longrightarrow \overline{%
b}=\frac b{{\cal N}}$ (the second Landau's parameter per lattice site)).
Thus, we obtain

\begin{equation}
\frac{\Delta c_{V,n}}{T_s{\left[ \Delta \left( \frac{\partial {\Delta }_s^2}{%
\partial T}\right) _{V,n,T=T_s}\right] }^2}=\overline{b}
\end{equation}
and

\begin{equation}
\frac{\Delta \left( \frac{\partial \mu }{\partial T}\right) _{V,n,T=T_s}}{%
\left( \frac{\partial T_s}{\partial n}\right) _V{\left[ \Delta \left( \frac{%
\partial {\Delta }_s^2}{\partial T}\right) _{V,n,T=T_s}\right] }^2}=%
\overline{b}.
\end{equation}
The specific heat per lattice site can be written in the form

\begin{equation}
c_{V,n}\left( T\right) =A\left( T\right) +B\left( T\right) {\left( \frac{%
\partial \mu }{\partial T}\right) }_{V,n}.
\end{equation}
The results of our calculations are presented in Fig. 1. We see (Fig. 1a, c)
that $\displaystyle{A\left( T\right) }$ and the temperature derivative of
the chemical potential exhibit jumps at $T=T_s$ whereas the coefficient $%
B\left( T\right) $ is a continuous function at the critical point (Fig. 1b).
It results in the jump of the specific heat per lattice site $c_{V,n}$ (Fig.
1d) at $T=T_s$. It would be interseting to know whether the temperature
behaviour of the coefficients (Fig.1a, b, c) in (73) is a general property
of all superconductors (when write the specific heat in the form (73)) or
only a property of this particular model. This question needs a verification
from the experimental (nobody has proved it till now) and also from the
theoretical point of view (when consider another microscopic models). From
the other hand, the model we have applied here, should qualitatively
reproduce all general properties of superconductors. Therefore it is very
probably that only coefficient $A$ and the temperature derivative of the
chemical potential exibit jumps at the critical point whereas the
coefficient $B$ is a continuous function for these materials. The plots of
the derivatives $\displaystyle{{(\frac{{\partial T}_s}{\partial n})}_V}$ and
$\displaystyle{{(\frac{{\partial \Delta }_s^2}{\partial T})}_{V,n}}$ are
given in Fig. 1e and Fig. 1f, respectively. Calculating the lhs of (71) and
(72) independently by means of the data displayed in the Figs 1c-f we obtain
almost the same value of the second Landau's parameter per lattice site $%
\overline{b}\approx $ $8.6\ eV^{-3}$. It also means that the formula (46) is
automatically fulfilled (see Fig. 1c, d, e).

\paragraph{Conclusions}

The general, phenomenological relations (16), (22) and (25)-(28) together
with their analogs (40)-(45) seem to be very important for superconducting
systems in the practice because they connect different macroscopic
thermodynamical quantities at the critical point. They also allowed to
obtain interesting cross-relations (29)-(36) and (46)-(53) between different
critical jumps and the critical temperature derivatives or between the
separate jumps. Due to a variety of relations we have a possibility to
estimate in an indirect way many of the critical jumps without performing
experiments. Such a possibility can especially be interesting in many cases
where experiments are difficult to perform and the information gained in
this way can be decisive as e.g. the indication for the hole-type
superconductivity in the high-$T_{c\text{ }}$superconductor $%
YBa_2Cu_3O_{7-\delta }$ (see Refs [6], [7] for details).

\paragraph{Note added in proof}
\newcounter{rown}[equation]
\setcounter{equation}{0}
\renewcommand{\theequation}{N. \arabic{equation}}
The derived critical relations (29)-(36) and (46)-(53) are also valid for a more general case of superconducting,
correlated systems with many order parameters (multiband systems).
As a consequence of electronic correlations the critical temperature $T_s$ is the same
for all order parameters (all of them vanish for $T \geq T_s$).
To make a proof of the statement that the relations (29)-(36) and (46)-(53)
are valid in this case we consider the Gibbs potential in the form
\begin{equation}
Z(T,p,N,x_1,x_ 2,...,x_n)=Z_0(T,p,N)+\sum\limits_{i=1}^{n}a_ix_i+\frac{1}{2}\sum\limits_{i,j=1}^{n}b_{ij}x_ix_j
\end{equation}
as a generalization of the Gibbs potential (4) after Taylor expansion with respect to the order parameters
$x_i$ ($i=1,2,...,n$). The parameters $a_i$, $b_{ij}=b_{ji}$ ($i,j=1,2,...,n$) have similar meaning as in the formula (4). Using the orthogonal transformation
\begin{equation}
x_i=\sum\limits_{j=1}^{n}{\alpha}_{ij}{\bar x}_j
\end{equation}
the bilinear form in (N.1) can always be diagonalized ($b_{ij}=b_{ji}$) and instead of the formulae (N.1) we obtain the following expression (similar to (4))
\begin{equation}
Z(T,p,N,{\bar x}_1,{\bar x}_2,...,{\bar x}_n)=Z_0(T,p,N)+\sum\limits_{i=1}^n {\bar a}_i{\bar x}_i+\frac{1}{2}\sum\limits_{i=1}^{n} {\bar b}_i{\bar x}_i^2
\end{equation}
where ${\bar a}_i=\sum\limits_{j=1}^{n}a_j{\alpha}_{ji}$ and ${\bar b}_i$ ($i=1,2,...,n$) are the eigenvalues of the matrix $b_{ij}$. The quantities ${\bar x}_i$ ($i=1,2,...,n$) can be treated as "new"
order parameters because of the reciprocal relation
\begin{equation}
{\bar x}_i=\sum\limits_{j=1}^n x_j{\alpha}_{ji}
\end{equation}
and due to the fact (see (N.4)) that ${\bar x}_{i}$ ($i=1,...,n$) vanish at the same critical temperature $T_s$
as the order parameters $x_i$ ($i=1,...,n$). From the minimum principle ($\frac{\partial Z}{\partial {\bar x}_i}=0$, $i=1,...,n$) we obtain
\begin{equation}
{\bar x}_i=-\frac{{\bar a}_i}{{\bar b}_i}
\end{equation}
and this equation can be used to eliminate ${\bar b}_i$ ($i=1,...,n$) from the expression (N.3). We obtain in this way very similar expression to (9), it is
\begin{equation}
 Z(T,p,N,{\bar x}_1,{\bar x}_2,...,{\bar x}_n)=Z_0(T,p,N)+\frac{1}{2}\sum\limits_{i=1}^n {\bar a}_i {\bar x}_i^2.
\end{equation}
To each ${\bar a}_i$ and ${\bar x}_i$ ($i=1,...,n$) we apply analogous assumptions to (7) and (8) where the lower index $i$ should be added. Repeating the calculations similar to the described above (see (10)-(28))
we obtain analogous expressions to (16), (22) and (25)-(28). These new expressions, however, can easily be obtained from (16), (22) and (25)-(28) when replace $b[\Delta(\frac{\partial x}{\partial T})_{p,N,T=T_s}]^2$
on the rhs of them by a new quantity $\sum\limits_{i=1}^n {\bar b}_i[\Delta(\frac{\partial {\bar x}_i}{\partial T})_{p,N,T=T_s}]^2$. This new quantity is irrelevant when forming the quotients (see (29)-(33)) because it cancels out.
Thus, the formulae (29)-(33) and the resulting cross relations (34)-(36) remain unchanged. Exactly the same can be said about the formulae (46)-(53) which remain also the same when apply the procedure described in this section with
respect to the Helmholtz potential ${\bar Z}={\bar Z}(T,V,N,x_1,x_2,...,x_n)$ as a generalization of (39) for the case of superconducting systems with many order parameters.

\end{document}